\documentclass{article}

\usepackage[utf8]{inputenc}
\usepackage{graphicx}
\usepackage{amsfonts}
\usepackage{tikz}
\usetikzlibrary{shapes,arrows}
\usepackage[shortlabels]{enumitem}
\usepackage{mathtools}
\usepackage{comment}
\usepackage{url}
\usepackage{tabularx}
\usepackage{tabu}
\usepackage{doi}
\usepackage[title]{appendix}

\usepackage{natbib}
\bibpunct{(}{)}{;}{a}{,}{,}

\usepackage{hyperref}
\usepackage{cleveref}

\title{Robust decision analysis under severe uncertainty and ambiguous tradeoffs: an invasive species case study}

\author{
  Ullrika Sahlin \\
  Centre of Environmental and Climate Sciences \\
  Lund University, Sweden \\
  \texttt{Ullrika.Sahlin@cec.lu.se}
  \and
  Matthias C. M. Troffaes \\
  Department of Mathematical Sciences \\
  Durham University, UK \\
  \texttt{matthias.troffaes@durham.ac.uk}
 \and
 Lennart Edsman \\
 Swedish University of Agricultural Sciences \\
 Department of Aquatic Resources, Sweden \\
 \texttt{lennart.edsman@slu.se}
}

\date{}

\newcommand{\distbeta}{\mathrm{Beta}}
\newcommand{\distbin}{\mathrm{Bin}}

\newcommand{\rewards}{\mathcal{R}}
\newcommand{\attrib}{\mathcal{A}}

\newcommand{\utilset}{\mathcal{U}}

\begin{document}

\maketitle

\paragraph{ABSTRACT}

Bayesian decision analysis is a useful method for risk management decisions, but is limited in its ability to consider severe uncertainty in knowledge, and value ambiguity in management objectives. We study the use of robust Bayesian decision analysis to handle problems where one or both of these issues arise. The robust Bayesian approach models severe uncertainty through bounds on probability distributions, and value ambiguity through bounds on utility functions. To incorporate data, standard Bayesian updating is applied on the entire set of distributions. To elicit our expert's utility representing the value of different management objectives, we use a modified version of the swing weighting procedure that can cope with severe value ambiguity. We demonstrate these methods on an environmental management problem to eradicate an alien invasive marmorkrebs recently discovered in Sweden, which needed a rapid response despite substantial knowledge gaps if the species was still present (i.e. severe uncertainty) and the need for difficult tradeoffs and competing interests (i.e. value ambiguity). We identify that the decision alternatives to drain the system and remove individuals in combination with dredging and sieving with or without a degradable biocide, or increasing pH, are consistently bad under the entire range of probability and utility bounds. This case study shows how robust Bayesian decision analysis provides a transparent methodology for integrating information in risk management problems where little data are available and/or where the tradeoffs ambiguous.




\section{INTRODUCTION}

\subsection{Uncertainty in Environmental Management}

Environmental risk managers must often make decisions under uncertainty, especially under multiple objectives \citep{NAP12568}. To take this uncertainty into consideration, uncertainty must be characterized, assessed and conveyed \citep{2014:fischhoff}. There are several types and sources of uncertainty to consider in management of environmental systems \citep{2011:maxim}, including ambiguity in the decision maker's objectives \citep{2008:ascough,2005:dewulf,2014:mccarthy}. 

Decision theory offers solutions on how to deal with uncertainty. However, these solutions must be transferred to practical applications, and explained in a way that enables managers to identify which solution to use for a particular problem. In addition to uncertainty due to lack of knowledge, eliciting judgments or preferences is sensitive to psychological factors resulting in cognitive biases \citep{OHagan_2006, hemming_burgman_hanea_mcbride_wintle_2017}. For example, value ambiguity can be stronger when uncertainty in outcomes is high. A structured approach to decision making is required to overcome cognitive biases in expert's judgments and values, and to make appropriate use of the relevant decision theory \citep{2012sdm,NAP12568}.

Bayesian decision analysis includes both learning and optimization, and is widely used in environmental management \citep{2012:carriger,2014:mccarthy,2016:carriger}. Bayesian analysis (statistical inference) for learning has been applied in environmental management to quantify uncertainty in management outcomes due to parameter uncertainty \citep{2013:heard, 2011:mcgowan,Hartig2018}, uncertainty in underlying mechanisms \citep{2012:buhle,VANOIJEN2013}, and to integrate expert knowledge \citep{hemming_burgman_hanea_mcbride_wintle_2017,Martin2012, OHagan_2006}.

In standard Bayesian decision analysis,
we start with a prior distribution over the parameters within the assessment model, which
embodies all expert information that is not captured in the data.
Next, we need a model to connect the data to the parameters.
The standard way of doing so goes via the likelihood function,
which models how the data are generated for any known fixed value of the
parameters.
Finally, we need a utility function which encapsulates the
decision maker's preferences over the possible decision outcomes.
The prior, likelihood, and utility function are then combined into
a so-called posterior expected utility.
The optimal decision is found by maximizing this quantity
over all decision alternatives.
In this way, Bayesian decision theory
combines prior knowledge with evidence, allows us to quantify uncertainty in the impact of decisions, and provides a method for selecting the optimal decision alternative. 

Standard Bayesian analysis is limited to uncertainty quantified by a single probability distribution.
A feature of Bayesian analysis is that when data are sparse, the
analysis hinges on a correct specification of the prior.
However, when experts cannot express their uncertainty with high confidence, or when they disagree, specifying a full prior distribution may be difficult.
In such cases, we would like to avoid a situation
where the analysis depends on arbitrary choices in our prior
\citep{1854:boole,1921:keynes,2007:troffaes:decision:intro, 2011:sahlinne}. 
A second issue is that the decision maker's preferences over outcomes
may only be partially quantified \citep{1962:aumann}, for instance due to the decision
maker's unfamiliarity with the outcomes, or due different interest
groups having conflicting goals (e.g. different relative values of management costs versus negative ecological impacts).

Therefore, some argue for alternative or second order expressions of uncertainty to handle situations where probabilities or utilities are not well known \citep{2014:fischhoff}.
For example, \citet{regan2005} proposed the use of information gap theory to deal with severe uncertainty in environmental management decisions, \citet{todd1998:fuzzy} proposed fuzzy sets to represent severe uncertainty about conservation status of species, and \citet{Lempert2007} addressed uncertainty through scenarios.  However, both assessors and decision makers may find it difficult to deal with alternative ways to express uncertainty, especially if it requires different methods for data analysis and modelling.

One way to resolve these issues is to work with sets of
prior distributions and sets of utility functions with respect to those aspects of the problem that cannot be
fully specified with confidence
\citep{1974:levi,1984:berger,1991:walley,1995:seidenfeld,2000:rios:bayesian:sens:anal}.
This allows us to work with weaker model assumptions.
For example, robust Bayesian analysis explores the
influence of the prior and the utility, through sensitivity analysis
on posterior inference \citep{2000:rios:bayesian:sens:anal}.
Specifically, robust Bayesian decision analysis performs a standard Bayesian analysis for each choice of prior and utility function in a given set.
The resulting bounds on the set of posterior expected utility values
can be interpreted as a
quantification of the decision maker's indeterminacy towards the
management decisions themselves resulting from lack of knowledge
\citep{1991:walley,2007:troffaes:decision:intro,2014:troffaes:itip:decision}.

In this paper, we study a way to deal with severe uncertainty both in values and in system knowledge by bounding probability distributions and utility functions. In this, what we refer to as, robust Bayesian decision analysis, we still express uncertainty using probability and utility, however we relax some of the requirements of standard Bayesian decision analysis. We demonstrate methods to learn from data and derive utilities by revisiting a real and typical environment management problem facing both severe uncertainty and value ambiguity.

To incorporate data, standard Bayesian updating is applied on a set of distributions. We allow for value ambiguity when eliciting utilities for different management alternatives, and there is a simultaneous propagation of imprecision in probability and utility in the analysis. 
In this approach, we identify management decisions that are consistently bad under the entire range of probability and utility bounds, and that we can therefore clearly exclude.
We also investigate how the performance of the remaining decisions varies as a function of our beliefs about the world.
Thereby, we show how the proposed robust version of Bayesian decision analysis enables a transparent use of information in environmental problems where little data are available and/or where the objectives are ambiguous.

\section{BACKGROUND}
\subsection{An Environmental Risk Management Problem}

In November 2012, specimens of the crayfish marmorkrebs \textit{Procambarus fallax forma virginalis}
were found in Sweden \citep{2013:bohman} from which twelve were instantly removed. The 2012-2013 winter was very cold, which may have reduced the chances of
any remaining individuals to survive. 
Marmorkrebs is an non-indigenous invasive species that recently has established in Europe \citep{2012:chucholl}. It reproduces by cloning \citep{2009:jones}, and therefore marmorkrebs constitute a high risk of bringing new disease vectors, and competition with native crayfish which are already threatened. 

According to the Swedish Species Protection Ordinance (2007:845), the species is forbidden to import, move and hold in Sweden. However, illegal activities occur and marmorkrebs could be released into the wild. 

In spring 2013, environmental managers were concerned that marmorkreb might still be present. A fast response enhances the chances of successful eradication, and therefore, despite uncertainty in the current state, decision making was urgent. 
A group of experts were assigned the task to evaluate the probability of crayfish presence and, together with stakeholders, perform a decision analysis to identify an appropriate action \citep{2013:bohman}.
Some disagreement sustained on how to balance costs, environmental impact, and efficiency. Indeed, whilst a radical decision has higher chances of eradication, it typically comes at higher societal and environmental cost. In addition, there was also uncertainty about the possibility for marmorkreb to successfully survive under each of the management options.

Meanwhile, a sampling scheme was set up to reduce uncertainty by collecting  evidence for the presence of marmorkreb. No marmorkreb were observed in any of the trials \citep{2013:bohman}. One conclusion might be that no marmorkreb is present, and therefore no action is needed. To do nothing was also the decision taken by managers in this particular case. A more reflective conclusion acknowledges that even though none were observed, the species, or pathogens brought in by the species, could still be there and actions might still be needed, especially when high values are at stake. 
Since 2013, there has been no further individuals observed in Sweden and no major outbreak of a disease associated with marmorkreb. 

\subsection{Decision Problems with Uncertainty, Value Ambiguity, or Both}

Environmental decision makers responsible for solving the marmorkreb problem described above face uncertainty in their knowledge about the system, as well as ambiguity in their values. System knowledge is the knowledge about the physical system and
the way in which we can interact with it, that is available for the environmental manager at the time for the decision. 
In this case, the lack of observed markmorkrebs in the summer following the introduction does not remove the need to evaluate management alternatives. Instead, the decision must be made under severe (or deep as in \citeauthor{NAP12568} (\citeyear{NAP12568})) uncertainty. 

Values influence the way a manager perceive and weight the outcomes of the alternative management actions. Combining severe uncertainty and value ambiguity, \citet{2011:sahlinne} identified four types of situations for decision making with respect to the clarity on uncertainty and values: 
\begin{itemize}
\item
In Type 1 situations, the decisions maker has extensive knowledge and information, expressed in terms of precise probability estimates. She also has clear and distinct preferences and values. 
\item
In Type 2 situations, the quality and quantity of information is poor, and it is difficult to represent the underlying uncertainty in terms of probability. On the other hand, the decision maker still has clear and distinct preferences and values: she knows what she wants and desires. 
\item
In Type 3 situations, the quality and quantity of information is good enough to assess precise probabilities. However, the decision maker lacks harmonious, clear and distinct preferences and values.
\item
In Type 4 situations, both information and preferences are unreliable or ambiguous.
\end{itemize}

Type 1 situations can be seen as the standard, where the usual principles for inference and reasoning work well. As long as you feel you have support to characterise uncertainty by subjective probability you are in Type 1 or Type 3. If not, you are in Type 2 or Type 4. Type 4 situations are not that uncommon. They arise for instance when the actions needed to prevent harm create a conflict within us, since we have to make difficult tradeoffs, and we may be unsure if there is a potential harm in the first place. In invasive species management, this could be rapid action to contain a potentially harmful species by killing all possible hosts within a distance from the sight of observation. Such rapid action creates a conflict between the ambition to protect e.g. trees (which can act as hosts to the alien species) and to eradicate the alien species \citep{2015:porth}. We as humans, are not very good in making such tradeoffs, especially when we are uncertain as well.

\subsection{Extending Bayesian Decision Analysis}

Bayesian decision analysis is able to solve Type 1 problems, while Type 2, 3 and 4 problems requires other ways to represent the impact of knowledge-based uncertainty and/or value ambiguity on the decision objectives and rules how to choose between management alternatives under uncertainty or value ambiguity. Robust Bayesian decision analysis is, as described in this paper, one way to simultaneously deal with uncertainty and value ambiguity. In addition, this approach enable a smooth transition of the specification of the decision analysis over all four types. 

In the next section we treat the marmorkreb management problem as a Type 4 problem (however the reasoning is the same under Type 2 and 3) and assess the probability of presence taking into account available evidence (i.e. by going from a prior to a posterior probability of presence). Uncertainty coming from limited knowledge about the system and ambiguous values is characterized by lower and upper bounds on the probability that the crayfish is present after management, and by lower and upper bounds on the expected utility.  

\section{METHODS}

\subsection{Management Alternatives}

The management problem is to seek the best management decision for eradicating
any alien crayfish possibly still in the water. To do so, the decision maker needs to assess the
probability of eradication across different decisions, as well as
the associated costs and environmental impacts.
In this particular case, the following management decisions were identified \citep{2013:bohman}: 
\begin{enumerate}[I,nosep]
\item Do nothing and inform the public about the problem with non-indigenous species and the need to prevent introductions.
\item Mechanical removal of individual specimens found by fishing.
\item Drain the system on water and removal of individuals by hand.
\item Drain the system of water, dredge and sieve the masses to identify and remove individuals.
\item Use a degradable biocide in combination with drainage to increase the biocide concentration.
\item Increase pH in combination with drainage and removal by hand.
\end{enumerate}

The decision problem is specified through a model that links the variables of the system state and the decision maker's values to parameters and data (\cref{fig:bhm}). The variables and dependencies of this probabilistic network (or, more precisely, influence diagram, since it also includes decision and utility nodes) is further explained in the next two sections.

\subsection{Model of the System}

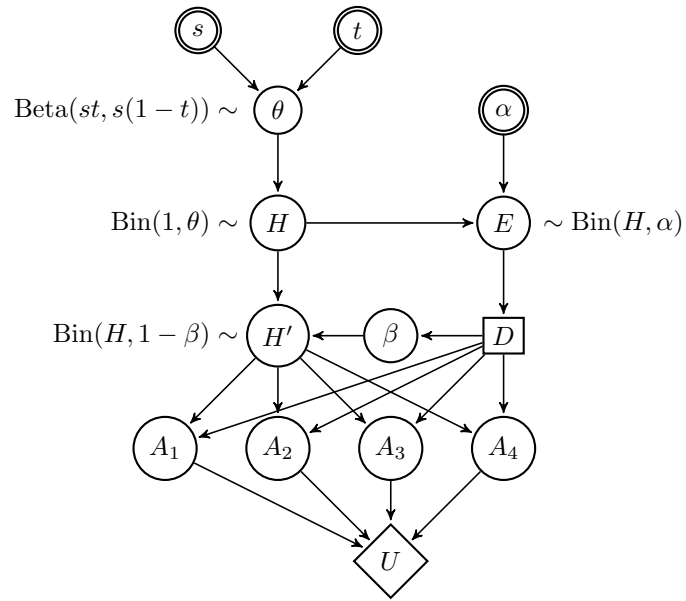
\begin{figure}
\begin{center}
\begin{tikzpicture}[->,>=stealth',shorten >=1pt,auto,node distance=1.5cm,
semithick]
\tikzstyle{const}=[fill=white,shape=circle,double,draw=black,thick,text=black];
\tikzstyle{var}=[fill=white,shape=circle,draw=black,thick,text=black];
\tikzstyle{dec}=[fill=white,shape=rectangle,draw=black,thick,text=black];
\tikzstyle{util}=[fill=white,shape=diamond,draw=black,thick,text=black];
\node[dec] (deci) {$D$};
\node[var] (beta) [left of=deci] {$\beta$};
\node[var] (hyp2) [left of=beta] {$H'$};
\node[var] (hyp) [above of=hyp2]  {$H$};
\node[var] (theta) [above of=hyp] {$\theta$};
\node[const] (s) [above left of=theta]{$s$};
\node[const] (t) [above right of=theta] {$t$};
\node[var] (evi) [above of=deci]  {$E$};
\node[const] (alpha) [above of=evi] {$\alpha$};
\node[var] (reward3) [below of=beta] {$A_3$};
\node[var] (reward2) [left of=reward3] {$A_2$};
\node[var] (reward1) [left of=reward2] {$A_1$};
\node[var] (reward4) [right of=reward3] {$A_4$};
\node[util] (util) [below of=reward3] {$U$};
\path (s) edge (theta);
\path (t) edge (theta);
\path (theta) edge (hyp);
\path (hyp) edge (evi);
\path (alpha) edge (evi);
\path (evi) edge (deci);
\path (deci) edge (beta);
\path (beta) edge (hyp2);
\path (hyp) edge (hyp2);
\path (hyp2) edge (reward1);
\path (deci) edge (reward1);
\path (reward1) edge (util);
\path (hyp2) edge (reward2);
\path (deci) edge (reward2);
\path (reward2) edge (util);
\path (hyp2) edge (reward3);
\path (deci) edge (reward3);
\path (reward3) edge (util);
\path (hyp2) edge (reward4);
\path (deci) edge (reward4);
\path (reward4) edge (util);
\node [node distance=0.4cm,left of=theta,left] {$\distbeta(st,s(1-t))\sim$};
\node [node distance=0.4cm,left of=hyp,left] {$\distbin(1,\theta)\sim$};
\node [node distance=0.4cm,left of=hyp2,left] {$\distbin(H,1-\beta)\sim$};
\node [node distance=0.4cm,right of=evi,right] {$\sim\distbin(H,\alpha)$};
\end{tikzpicture}
\end{center}
\caption{The graphical model for the crayfish management problem is a probabilistic network of nodes constituting of system states ($H$ presence before management, $H'$ presence after management), evidence (data) ($E$), parameters ($\theta$, $\beta$), hyperparameters ($s$, $t$, $\alpha$), decision ($D$), attributes (Biotic impact ($A_1$), Longevity of impacts ($A_2$), Feasibility ($A_3$) and Cost ($A_4$)) and utility ($U$).
\label{fig:bhm}}
\end{figure}

The state variable $H$ describes whether the crayfish is present in the system ($H=1$) or not ($H=0$). As we are uncertain about
the value of $H$ after the winter, we introduce a parameter $\theta$ which embodies the
probability that $H=1$, i.e. that the alien crayfish is still present during spring 2013. Following
the standard Bayesian approach, to allow us to learn about $\theta$
from data, we need to express our initial belief on this probability by a prior distribution on $\theta$. In this case, we
assume that $\theta$ follows a $\distbeta(st,s(t-1))$ distribution,
where $s$ and $t$ are hyperparameters satisfying $s>0$ and $0<t<1$.
We use Walley's parametrisation \citep[Sec.~7.7.3]{1991:walley}
to allow for a straightforward
interpretation of the parameters: $t$ is the prior expectation of
$\theta$, and $s$ controls the variance of the prior
(larger $s$ corresponding to smaller variance).

We learn about $\theta$ through the empirical evidence $E$, where $E=1$ if the
crayfish has been observed in the trial fishing during the summer, and $E=0$ otherwise. If
no alien crayfish is present, obviously none will be
observed. However, even if alien crayfish is present, we are only able to detect it with probability $\alpha$. It is crucial to consider such observation errors when learning from data. In this case, the detection probability reflects that the system is only partially observable.

The presence of the crayfish after the management has taken place is expressed in the model by $H'$. Efficacy, measured as the probability of successful eradication ($H'=0$), is captured by the parameter $\beta$. Since different management methods have different chances of success, $\beta$ depends on the management decision $D$. When decisions impact probabilities, we say that we have \emph{act-state dependence}. This will be important later when we perform sensitivity analysis.

Bringing everything together into a probabilistic causal network (\cref{fig:bhm}), the future state of the system is linked to the decision node and the the evidence is linked to the current state of the system. The laws of probability and Bayesian updating allow us to revise the probability of future state $H'$ given evidence $E$ and decision $D$. This is Bayesian learning, prediction and reasoning.    

In this problem, the experts (the assessors) were quite uncertain about the probability $\theta$ of the crayfish being present, about the chance $\alpha$ to see it in the trial fishing,
and about the efficacy (the probability $\beta(d)$) of eradication.
To consider these uncertainties, we could choose to could put a
subjective probability distribution over these parameters. When to use probability or not is a matter of choice. In this case, our uncertainty about these parameters is severe and we do not know which probability distribution to use. Instead, we choose to reflect this uncertainty using robust Bayesian analysis. 
Since we can learn about $\theta$ via $E$, we choose to model $\theta$ via a set of Beta distributions, with prior mean $t\in[0.1,0.9]$.
We cannot learn about $\alpha$ and $\beta(d)$, so to keep the analysis simple, we simply model these with intervals,
covering the full range of values that we might reasonably expect.
The experts stated that $\alpha\in[0.1,0.5]$ is a plausible range for detection probabilities in trial fishing of crayfish. 

Experts were asked to elicit $\beta(d)$, the efficacy
parameter in our model, representing the probability that the management alternative $d$ is
successful in eradication.
The resulting bounds on the efficacy $\beta(d)$ of successful eradication under the different management decisions $d$ are given in \cref{tab:beta}. A biocide in combination with drainage (V) was judged to always result in successful eradication. Increase pH in combination with drainage and removal by hand (VI) was judged as the second most efficient intervention. The option to drain the system of water, dredge and sieve the masses and remove individuals (IV) were judged to potentially be more successful that the option to drain the system on water and removal of individuals (III), but the experts were uncertain and could not clearly say which one was better than the other. Mechanical removal of individuals had the lowest probability of success. 
\begin{table}
\caption{Lower and upper bounds on the probability of successful eradication $\beta(d):=P(H'=0|d)$ for different decisions $d$.
\label{tab:beta}}
\begin{center}
\small
\begin{tabular}{l||cccccc}
&\multicolumn{6}{|c}{\textbf{Decision $d$}}\\
\textbf{Probability} &\textbf{I} & \textbf{II}& \textbf{III}& \textbf{IV} & \textbf{V}& \textbf{VI}\\
\hline
$\underline{\beta}(d)$&0&0.05&0.3&0.4&1.0&0.7 \\
$\overline{\beta}(d)$&0&0.25&0.5&0.7&1.0&0.8
\end{tabular}
\end{center}
\end{table}

\subsection{Derivation of Utility}
To elicit our expert's utility representing the value of different management objectives, we use a modified version of the swing weighting procedure that can cope with severe value ambiguity. A simplified version of the marmorkrebs problem was already treated in \citet{2017:troffaes::swing}, which focused on the theoretical results behind the swing weighting method that we will also use here. In this current paper, we treat the modelling of the likelihood in far more detail, we elicit utility from an expert, and we also focus on the simultaneous propagation of imprecision in probability and utility in a much more realistic setting. It is also possible to use judgements from several experts, but that is beyond the scope of this study.

The impact of the overall outcome for the management problem is described by various attributes identified
as relevant by a group of experts and stakeholders. These were biotic effects and longevity of impacts, feasibility and cost of the method. 
These attributes, denoted by $A_1$, \dots, $A_4$, are influenced both by the decision ($D$) and by
whether crayfish is still present or not ($H'$) (\cref{fig:bhm}).
The decision is evaluated through a joint utility function $U$
on these attributes. 

Each management decision was scored according to the attributes (\cref{tab:scores1}) using a Likert scale ranging from 1 to 4 constructed for each of these attributes, where
1 corresponds to the worst outcome, and 4 corresponds to the best
outcome (detailed descriptions of all attribute levels are in Appendix \ref{app:scores}).
The expert assessed attribute scores in case of successful eradication for each management decision (\cref{tab:scores1}). The expert, with more than 30 years of experience in crayfish management, based these scores on a literature review on techniques to eradicate freshwater crayfish \citep{2013:bohman}.
Here, the expert provided a point score for every management alternative. We note that, in our analysis, it would also have been possible to assess these scores using a range.
In case of failure to eradicate the invasive species ($H'=1$), the scores for biotic impact and longevity of
impacts drop to their worst values (i.e. a score of 1).  

\begin{table}
\caption{Scores (Likert scale 1 to 4) for each attribute and each management decision in case of a successful eradication of the crayfish.
\label{tab:scores1}}
\begin{center}
\small
\begin{tabular}{l|l|l|cccccc}
&\textbf{Worst} & \textbf{Best} &\multicolumn{6}{|c}{\textbf{Decision $d$}}\\
\textbf{Attribute} & \textbf{(score 1)} & \textbf{(score 4)}&\textbf{I} & \textbf{II}& \textbf{III}& \textbf{IV} & \textbf{V}& \textbf{VI}\\
\hline
Biotic impact &High&Low&4&4&3&3&2&2\\
Longevity of impacts&Long&Short&4&4&3&3&2&1 \\
Feasibility&Difficult&Easy&4&4&3&2&1&2\\
Cost&High&Low&4&4&3&1&2&3
\end{tabular}
\end{center}
\end{table}

In order to combine the scoring on all attributes into a utility, we first interpret the scores in \cref{tab:scores1} as marginal utilities (i.e. $U_i(a_i)=a_i$) and make a structural assumption that the joint utility function is a weighted sum of the individual marginal utilities.
So, for given scores (below referred to as a joint reward) $r\coloneqq (a_1,\dots,a_n)$, we assume that:
\begin{equation}
  U(r)\coloneqq \sum_{i=1}^n k_i U_i(a_i).
\end{equation}

Although this additive form restricts quite substantially the type of preferences that can be expressed,
the attraction of the linearity assumption is that it reduces the elicitation of the joint utility
to just the elicitation of the weights $k_1$, \dots, $k_n$.
Relaxing this additive form is theoretically possible but
unfortunately it makes the multi-attribute elicitation problem far
more complicated, with many more parameters to be identified, and with
the joint utility function becoming a non-linear function of the
marginal utility functions, even under full mutual utility
independence \citep[Theorem~6.1]{1993:keeney::multiattribute}. For
simplicity, here, we will therefore assume an additive form.
Also, since the resulting utility functions are unique up to a positive linear transformation, we may impose all weights to sum to one. 
As a consequence, we only need to elicit $n-1$ of the weights. There are many ways to elicit weights, and we choose an indirect method since experts can find it difficult to interpret the weights directly.

\subsection{Modelling Ambiguity in Attribute Weights}

In order to deal with possible unclear objectives (Type 3 and 4 situations), we will use a method for indirect elicitation described in \citet{2017:troffaes::swing} which models ambiguity in attribute weights. The method allows an almost arbitrary set of rewards to be compared to match the expert's experience, which also allows for ambiguity in the way the different attributes are weighed.
For a detailed mathematical description of the method, we refer to Appendix \ref{app:swingmethod}.
The elicitation method is consistent
under fairly relaxed conditions, which are satisfied in the setting that we
shall study here \citep[Sec.~6]{2017:troffaes::swing}.
It also includes the well known swing weighting method \citep{1986:winterfeldt} as a special case.

A downside of swing weighting is that it considers rewards which
are unnatural for our specific problem, because they consider extreme combinations of attributes, with all but one in their worst state. Therefore, experts may find it difficult to express their preferences over these rewards. From an impact assessment perspective it would instead be more natural to compare rewards made up by only small changes from a reference state. These are thus easier to compare
(regardless of any imprecision in preferences). The method we use is a generalization of swing weighting which deals with these problems.

To simplify the elicitation, we developed a graphical user interface (R code in Supplementary material) using shiny R \citep{shinyr} where the expert goes through the steps annotated below. We ran the elicitation procedure with the same expert twice, and what is reported below is the second iteration. The first iteration had slightly different choices for levels, but the overall conclusions remained the same.
\begin{enumerate}
\item The expert is informed of all attributes, along with a detailed description of all attribute levels (see Appendix \ref{app:scores}).
The expert had some input in setting realistic outcomes for these levels. Throughout the interface, short textual descriptions are used for the levels, rather than numbers, to ensure clarity throughout.
\item The expert is informed that they will be asked to compare these attributes at two levels. As a first step, the expert is asked to identify which pairs of levels they find most comfortable with comparing. Note that, at this stage, we excluded the `no impact' outcomes (level 4) for biotic impact and longevity to ensure meaningful joint outcomes are compared for the next steps.

For example, the expert chose levels $\{1,2\}$ for biotic impact, $\{2,3\}$ for longevity, $\{1,3\}$ for feasibility, and $\{1,3\}$ for cost where the highest levels comprise the reference state.
From these levels, we construct the following joint rewards (directly expressed in terms of marginal utilities):
\begin{center}
\begin{tabular}{l}
rewards \\
\hline
$u_0\coloneqq(1, 3, 3, 3)$ \\
$u_1\coloneqq(2, 2, 3, 3)$ \\
$u_2\coloneqq(2, 3, 1, 3)$ \\
$u_3\coloneqq(2, 3, 3, 1)$ \\
$u_4\coloneqq(2, 3, 3, 3)$
\end{tabular}
\end{center}
Here, $u_4$ is the reference state, $u_0$ modifies $u_4$ in the first attribute,
\dots, and $u_3$ modifies $u_4$ in the fourth attribute.
\item The expert is asked which of the above joint rewards is the worst outcome. In our case, the expert chose $u_2$,
so $r_2\preceq r_j\preceq r_4$ for all $j\in\{0,1,3\}$ (the symbol $\preceq$ means `is less or equally preferred to').
\item Next, we introduce uncertainty in the rewards using lotteries. Given two rewards $a$ and $b$, and a number $\alpha\in[0,1]$, the expression
  \begin{equation}
    \alpha a\oplus (1-\alpha) b
  \end{equation}
  denotes an uncertain reward where $a$ is obtained with probability $\alpha$ and
  $b$ is obtained with probability $1-\alpha$. Comparing a lottery with a known reward is a common way to indirectly elicit someone's probability of a random event, or someone's utility of a reward.
 
\item
  Uncertainty in weights is obtained by asking the expert to compare and set values on $\alpha$ in a range of rewards (as prescribed in Appendix
  \ref{app:swingmethod}). Our expert arrived at:
\begin{align}
  \label{eq:expertprefs1}
  0.60 r_2 \oplus 0.40 r_4\preceq &r_0\preceq 0.35 r_2\oplus 0.65 r_4 \\
  0.50 r_2 \oplus 0.50 r_4\preceq &r_1\preceq 0.40 r_2\oplus 0.60 r_4 \\
  0.10 r_2 \oplus 0.90 r_4\preceq &r_3\preceq 0.04 r_2\oplus 0.96 r_4
\end{align}
For instance, \cref{eq:expertprefs1} means that the expert prefers the
certain outcome $r_0$ over the uncertain outcome where $r_2$ happens
with 60\% chance and $r_4$ with 40\% chance.  However, when the chance
for $r_2$ is reduced to 35\% and the chance for $r_4$ is increased to
65\%, the expert prefers the uncertain outcome instead.
The other preferences have a similar interpretation.
\end{enumerate}

These assessments then lead to a set of linear inequalities
that determine a convex set of attribute weights.
In this decision analysis it is enough to consider extreme points of this set
(see Appendix \ref{app:swingmethod}) to derive bounds on expected utility. The extreme points were here calculated using the double description method
\citep{1996:fukuda} through the rcdd package in R \citep{r} (R code in Supplementary material).

\subsection{Select Decision}

The last step in the decision analysis, after specifying the parameters of a model to express our beliefs and values of alternative outcomes, is to choose the best management alternative. Because the decision affects the probability of successful management (i.e. we have 
act-state dependence), we have to treat the problem using interval dominance
\citep{2007:troffaes:decision:intro}. If probabilities and utilities are precise (Type 1 situations), then this is equivalent to the conventional approach of maximizing expected utility.  
In interval dominance, we consider the posterior expected utility interval
of every option. The set of decisions whose intervals are undominated are
then considered as optimal. If there is only one such management alternative, then obviously
that is the decision we ought to pick.

With interval dominance, it is always the case that the best worst case option dominates all non-optimal options. To help visualize this, we depicted the best worst case utility as a vertical dashed line on all plots: every option whose interval intersects with this line is optimal.

If there are multiple undominated management alternatives,
then this means that we have insufficient information to say which is the best. If so, we can deselect poor alternatives and arrived at a set of possible alternatives to select from.
One might then try to refine the set by collecting more information and rerun the analysis,
pick the alternative with the best worst outcome,
or decide which alternative to pick based on other concerns.
It is also possible to refine the interval analysis and eliminate further options by performing a sensitivity analysis over parameters that are not affected by the decision. We defer a discussion of this to \cref{sec:results}.

We estimate the posteriors
for each decision alternative $d$,
each extreme value of $t$, $\alpha$ and $\beta(d)$,
and each extreme attribute weight vector $k$ from \cref{tab:extremeweights}.
The utility was then evaluated through
\begin{equation}
  U(d,k)\coloneqq H'\sum_{i=1}^5 k_i U_i(a_i(H'=1,d)) + (1-H')\sum_{i=1}^5 k_i U_i(a_i(H'=0,d))
\end{equation}
where the marginal utilities were taken from \cref{tab:scores1},
and the posterior distribution for $H'$ was sampled using the graphical
model depicted in \cref{fig:bhm} using MCMC sampling in the R package rjags calling JAGS \citep{Plummer03jags:a} (R code in Supplementary material). 
Because inferences for lower $s$ values lead to tighter inferences, fixing $s$ to any specific value automatically covers all lower values for $s$ as well \citep{1996:walley::idm}. Therefore, we need not consider intervals for $s$, and only need to consider a reasonable upper bound.
In our analysis, the parameter $s$ was set to 2. 
This choice ensures that all typical choices of precise Bayesian prior distributions for Bernoulli sampling are covered \citep{1996:walley::idm}.

\section{RESULTS}
\label{sec:results}

\begin{figure}
\begin{center}
\includegraphics[width=\textwidth]{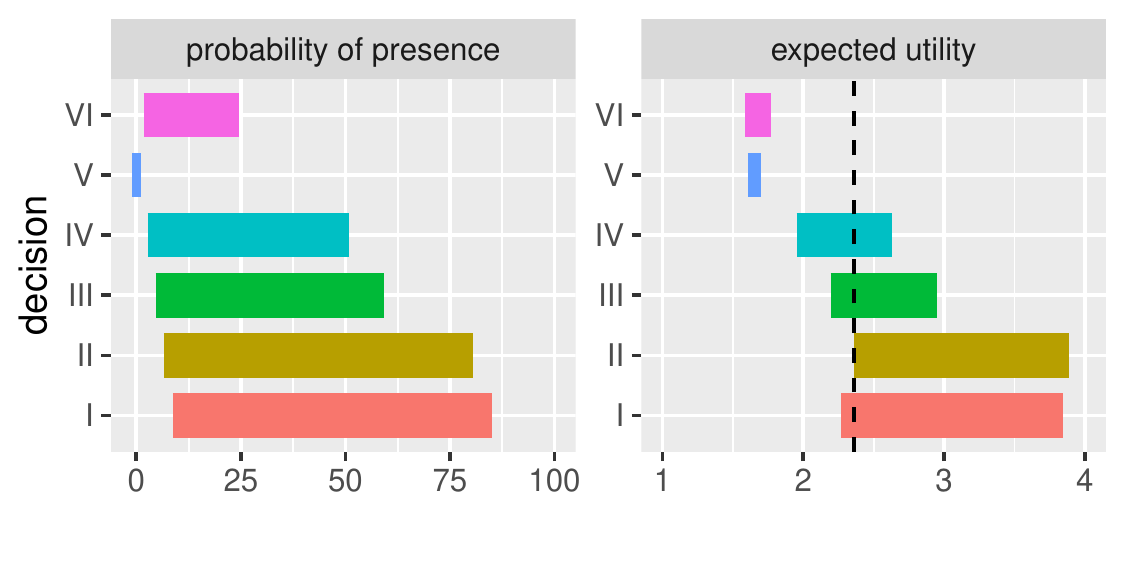}
\end{center}
\caption{Intervals for posterior probability of crayfish being present after management $P(H'=1|d)$ (left) and expected utility $E(U(d))$ (right) for different decision alternatives $d$: I. Do nothing, II. Mechanical removal by fishing, III. Drainage and removal by hand, IV. Drainage, dredging and sieving before removal by hand, V. Use a degradable biocide in combination with drainage, and VI. Increase pH in combination with drainage and removal by hand, given the prior probability $0.1 \leq t \leq 0.9$ 
, prior equivalent sample size $s=2$, and detection probability 
$0.1 \leq \alpha \leq 0.5$. The highest worst expected utility is indicated by a vertical dashed line. In this case decision alternatives V and VI are dominated.
\label{fig:results:all}}
\end{figure}

Interval dominance evaluated from the full robust Bayesian analysis reveal that the options ``Use a degradable biocide in combination with drainage`` (V) and ``Increase pH in combination with drainage and removal by hand`` (VI) are dominated by the other management alternatives (\cref{fig:results:all}). 
The intervals on the probability of presence after management are in this analysis a consequence from considering uncertainty in terms of sets of probability distributions for the parameter $\theta$ (arising from combinations of values for prior probability $t\in[0.1,0.9]$ and detection probability $\alpha\in[0.1,0.5]$) and intervals on the parameter $\beta(d)$ (the efficiency of each management alternative $d$ (\cref{tab:beta})). The expected utility intervals are a consequence from this uncertainty about the probability of presence after management and sets of values for the utilities.
The expected utility intervals on the four non-dominated decision alternatives in \cref{fig:results:all} are wide and partially overlapping. Therefore, they are all reasonable, but highly uncertain.

\begin{figure}
\begin{center}
\includegraphics[width=\textwidth]{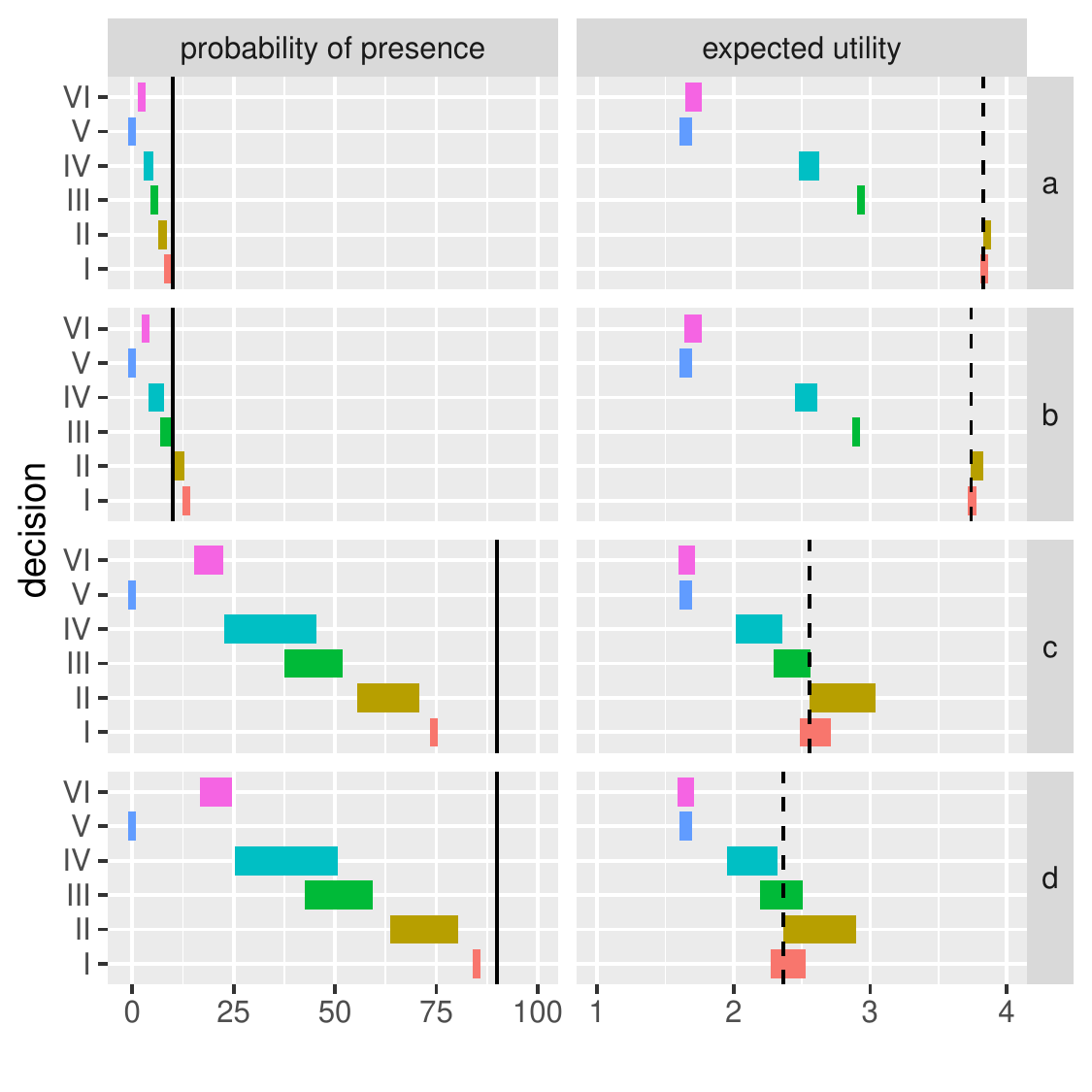}
\end{center}
  	\caption{Intervals for posterior probability of crayfish being present after management $P(H'=1|d)$ (left) and expected utility $E(U(d))$ (right) for different decision alternatives $d$, for each of the extreme points for prior expected probability (solid line) and detection probability a) $t=0.1,\alpha=0.5$, b) $t=0.1,\alpha=0.1$, c) $t=0.9,\alpha=0.5$, d) $t=0.9,\alpha=0.1$, and prior equivalent sample size $s=2$. The highest worst expected utility is indicated by a vertical dashed line. In this case decision alternatives IV, V and VI are dominated. \label{fig:results:extremes}}
\end{figure}

A refined analysis evaluating interval dominance for different beliefs in the system state and observation error, may reveal if there are any further dominated alternatives. Uncertainty about the efficiency of management and value ambiguity are not refined in this step, because they are specified in a way such that there is no straightforward way on how refine them any further.
We therefore study if different choices of the hyperparameters $t$ and $\alpha$ from the range reflecting our beliefs, result in additional conclusions about dominance (\cref{fig:results:extremes}). 

No individuals were observed in trial fishing, and therefore the probability of the crayfish being present is bounded from above by the prior probability of crayfish presence ($t$), as long as the the detection probability is large enough. As seen from the left hand side of \cref{fig:results:extremes}: the prior (solid vertical line) is almost always at least as high as the interval for the posterior probability of presence after management, with the exception of when $t=0.1$ and $\alpha=0.1$.

A higher prior belief of crayfish presence logically results in a high posterior probability, and in lower values for the expected utility. Also, the differences in expected utilities between decision alternatives becomes smaller, since the utility from the loss when the species is present gets a higher weight compared to the specific utility under each decision alternative. 
In contrast, a higher detection probability results in lower risk and narrower bounds for the posterior. This is expected since we put more trust to data compared to the prior belief, and since we do not observe any specimens, the posterior probability becomes relatively lower than before. 

For each extreme point of the hyperparameters, we find by the refined analysis, that the option ``Drainage and removal by hand`` (IV) is dominated as well (in addition to V and VI) (\cref{fig:results:extremes}). All three options ``inform only`` (I), ``mechanical removal'' (II), and ``drain the system on water and removal of individuals by hand'' (III), are reasonable. We also see that the option II is the best worst case decision since it has the highest value on the worst possible expected utility (\cref{fig:results:extremes}). The management alternative II also dominates all non-optimal options (since its lower bound on the expected utility exceeds the upper bounds on the expected utilities of IV, V and VI). 

\section{DISCUSSION}

We demonstrated the use of robust Bayesian decision analysis to solve
a risk management problem under severe uncertainty and value ambiguity.
Bayesian decision theory is applicable in situations when epistemic uncertainty is judged to be reliably characterized by subjective probability and preferences and values are clear and distinct (a Type 1 problem according to \citet{2011:sahlinne}). Decision problems may face uncertainty in knowledge bases (Type 2), ambiguity in values (Type 3) or both (Type 4). Type 3 problems can occur when there is ambiguity in values from multiple actors and different frames \citep{2005:dewulf}. Here we describe the decision to eradicate an invasive alien species to be taken under severe uncertainty and value ambiguity (Type 4). 

To handle the limited possibility to verify the presence of the alien species,
the small chances in detecting the species if present,
and difficulties in predicting the probability of successful eradication under different management options,
we opted to model certain quantities through sets of probability distributions. There will always remain some degree of subjectivity in setting
probability bounds. However, this is no different from standard probability elicitation \citep{OHagan_2006, hemming_burgman_hanea_mcbride_wintle_2017}. 
Similarly, to handle the severe ambiguity in how decision makers weigh the different attributes
in the outcome (e.g. cost, feasibility, biotic impact),
we used an extension of the standard swing weighting method
to model these ambiguities through sets of utility functions.
We then propagated these entire sets through a robust Bayesian analysis, and then compared
the posterior expected utility intervals to identify the best possible decisions using interval dominance.

Here, robust Bayesian decision analysis is presented as a modification of Bayesian decision analysis, where the ability to incorporate prior beliefs and evidence is ensured, but uncertainty is treated in a more conservative way and ambiguity in values are acknowledged. This decision theory includes learning under severe uncertainty to identify management decisions that are consistently bad under a plausible range of probability and utility bounds, and that decision makers can therefore clearly exclude. Since updating and decision analysis is done in a single process, it is easy to evaluate sensitivity towards the initial choice of our beliefs about the world. 

For the marmorkreb problem, we found that three out of six management alternatives were non-dominated. This conclusion was found after a refined analysis, where interval dominance was evaluated within choices of hyperparameters (similar to paired testing). The analysis gives support to the decision that actually was taken, i.e. to do nothing, but it also shows that mechanic removal by fishing would have been a better choice. 

The problem to choose if, and how, to eradicate the alien invasive crayfish is a simple one (from a structural point of view). Note that there are several examples of somewhat similar Bayesian decision analyses to support management of invasive species \citep{2017:Russell, 2017:sakamoto, 2014:rout, 2015:Clarke, Regan2011}. These works address the question as to how long we need to monitor after an eradication attempt to be certain the species is gone. In contrast, in this paper, the management decision is if eradication should be done, and if so, which way to do it. 

The calculations for the analysis in this paper can be done within seconds (the R code is available as Supplementary material). For larger problems, with more complex structures, the computational challenge of Bayesian updating and optimization under sets of distributions quickly becomes resource demanding. Future applications of robust Bayesian analysis on risk management require efficient algorithms for learning e.g. relying on MCMC sampling for any type of model or approximations for specific types of models \citep{Rue_INLA_2017}. Today, those algorithms exist for standard Bayesian analysis (e.g. \citet{Plummer03jags:a}).

Value ambiguity may be a larger concern than uncertainty in risk management problems. Eliciting the set of prior distributions and set of utility functions poses a practical challenge. 
Although the extended swing weighting method was chosen for its strong
consistency properties, it's not clear whether
such elicitation would work in a practical setting.
In our elicitation procedure, we asked the minimal number of questions to elicit weights, similar to the standard swing weighting method. In order to check the internal consistency, one could elicit further preferences and verify that these are compatible. If inconsistencies appear, our suggestion is to communicate these inconsistencies back to the expert, and ask them to reconsider their preferences to achieve consistency.

In the shiny R app, the expert is given some information about what constitutes a lottery, and how to compare lotteries, involving some hypothetical rewards. We found that this substantially helped the expert to conduct the next step, although it still remained a conceptual challenge for someone not experienced in utility elicitation. Therefore, we recommend the procedure to be run with a facilitator who has good conceptual understanding of the procedure.

We note that robust Bayesian decision analysis is not limited to the how utility was derived in this example.
Moving away from an additive utility function is possible at the expense of a more complicated elicitation problem, where the joint utility function becomes a non-linear function of the marginal utility functions \cite{1993:keeney::multiattribute}.
We chose an additive utility function to keep the analysis straightforward and to allow for linear optimisation, which makes it easy to perform the bounding computationally, even though this choice obviously restricts the type of preferences that can be expressed.
Additionally, we note that the analysis could be expanded to also account for uncertainty in the assessed consequences beyond just eradication and detection. Even though in this case study the expert did not reveal any uncertainty in his assessment, one might well imagine a scenario where under some of the decision alternatives, there is uncertainty in the biotic impact, longevity, feasibility, or cost. One could account for this through probability, or through probability bounding if this uncertainty is severe, making for a more advanced model and more complex analysis.

Any serious attempt to deal with uncertainty and value ambiguity ought to find transparent ways to adapt to the type of decision problem at hand (Type 1 to 4). We suggest that relaxing the assumptions behind standard Bayesian decision theory into robust Bayesian decision theory is one way to do this, and goes from one rigorous principle for learning and quantifying epistemic uncertainty into another \citep{1991:walley,2000:rios:bayesian:sens:anal}.        

\section{Acknowledgements}
US was supported by the Swedish research council FORMAS through the project ``Scaling up uncertain environmental evidence'' (219-2013-1271)
and the strategic research environment Biodiversity and Ecosystem Services in Changing Climate (BECC). LE was funded by the Swedish Agency for Marine and Water Management. 

\section{Authors’ contributions}
US and MT conceived the ideas and designed methodology; US and LE provided the case study; LE provided expert judgment; US and MT did the analysis with equal contribution and the writing of the manuscript; LE reviewed the manuscript for clarity; all have given final approval for publication.

\section{Supplementary material}

The R code for performing the analysis is found at \url{https://github.com/mcmtroffaes/r-crayfish-risk-analysis/}.

\bibliographystyle{apa} 
\bibliography{refs}

\clearpage

\pagebreak

\begin{appendices}

\section{Attribute tables}
\label{app:scores}

\Cref{tab:scale1,tab:scale2,tab:scale3,tab:scale4}
list and describe the scales for all attributes.

\begin{table}
\caption{Likert scales for the biotic impact attribute.\label{tab:scale1}}
\begin{center}
\small
\begin{tabu}{r|X|X[2]}
  level & short description & description \\ 
  \hline
  4 & no impact & No negative impacts. \\
  3 & minor impact on some & Some species are negatively affected, but this does not have any impact on the viability of their populations and the invasive alien species is not present in the system. \\ 
  2 & major impact on some & Some of the species in the system are negatively affected or that the majority of species are affected but not with any impact on the viability of their populations, and the invasive alien species is not present in the system. \\ 
  1 & major impact on most & Majority of the species in the system are negatively affected or the invasive alien species is present in the system.
\end{tabu}
\end{center}
\end{table}

\begin{table}
\caption{Likert scales for the longevity attribute.\label{tab:scale2}}
\begin{center}
\small
\begin{tabu}{r|X|X[2]}
  level & short description & description \\ 
  \hline
  4 & no impact & No negative impacts. \\
  3 & month & Duration of negative biotic impacts up to a month. \\ 
  2 & 1 year & Duration of negative biotic impacts up to 1 year. \\ 
  1 & $>$ 1 years & Duration of negative biotic impacts for more than 1 years.
\end{tabu}
\end{center}
\end{table}

\begin{table}
\caption{Likert scales for the feasibility attribute.\label{tab:scale3}}
\begin{center}
\small
\begin{tabu}{r|X|X[2]}
  level & short description & description \\ 
  \hline
  4 & no obstacles & No major obstacles in carrying out the method. \\ 
  3 & minor obstacles & Some obstacles to carry out the method, but these are possible to overcome in the current legislation and policy. \\ 
  2 & some controversy & Method is controversial and it requires a lot of preparatory work to be possible to carry out. \\ 
  1 & large controversy & Large controversy about the method and it may be in conflict with current legislation or policy. \\ 
\end{tabu}
\end{center}
\end{table}

\begin{table}
\caption{Likert scales for the cost attribute.\label{tab:scale4}}
\begin{center}
\small
\begin{tabu}{r|X|X[2]}
  level & short description & description \\ 
  \hline
  4 & $<$ 50k & Between 0 and 50 000 SEK. \\ 
  3 & 50-250k & Between 50 000 and 250 000 SEK. \\ 
  2 & 250-500k & Between 250 000 and 500 000 SEK. \\ 
  1 & $>$ 500k & More than 500 000 SEK.
\end{tabu}
\end{center}
\end{table}

\pagebreak

\section{Utility Elicitation: Mathematical Details}
\label{app:swingmethod}

In the setting of \citet{2017:troffaes::swing},
$\rewards\coloneqq\attrib_1\times\dots\times\attrib_n$ is a finite set of rewards,
each reward $r=(a_1,\dots,a_n)$ comprising of $n$ attributes.
A \emph{lottery} $\ell$ on $\rewards$ is a probability mass function over $\rewards$,
and is interpreted as a random reward with precisely known probabilities.
The set of all lotteries over $\rewards$ is denoted by $L(\rewards)$.
A \emph{utility function} on $\rewards$ is any function $U\colon\rewards\to\mathbb{R}$,
where we lift $U$ to $L(\rewards)$ in the usual way:
\begin{equation}
U(\ell)\coloneqq\sum_{r\in\rewards}\ell(r)U(r).
\end{equation}
We wish to model our preferences between lotteries over our multi-attribute rewards.
We will assume that our preferences
form a preorder $\succeq$ on $L(\rewards)$,
and can be represented through a \emph{set} $\utilset$
of utility functions $U\colon L(\rewards)\to\mathbb{R}$:
\begin{equation}
\ell_1\succeq\ell_2
\qquad\iff\qquad
\forall U\in\utilset\colon U(\ell_1)\ge U(\ell_2)
\end{equation}
for all $\ell_1$ and $\ell_2\in L(\rewards)$.
For theoretical foundations behind such representation,
we refer to \citet{2006:nau::incompleteprefs}.
Elicitation is then concerned with finding a procedure for identifying $\utilset$.

The elicitation method goes as follows \citep{2017:troffaes::swing}:
\begin{enumerate}[nosep]
\item Consider \emph{any} joint rewards $r_0$, \dots, $r_n$ such that for all $j\in\{1,\dots,n-1\}$
we have that
\begin{equation}
r_0\preceq r_j\preceq r_n
\end{equation}
\item For all $j\in\{1,\dots,n-1\}$, find the largest $\underline{\alpha}_j$
and smallest $\overline{\alpha}_j$ such that
\begin{equation}\label{eq:swingimprecise1}
(1-\underline{\alpha}_j)r_0 \oplus \underline{\alpha}_j r_n
\preceq r_j\preceq
(1-\overline{\alpha}_j)r_0 \oplus \overline{\alpha}_j r_n
\end{equation}
where $\oplus$ denotes the combination of rewards into lotteries,
so $(1-\alpha) r_1\oplus \alpha r_2$
is the lottery $\ell$ with $\ell(r_1)=1-\alpha$,
$\ell(r_2)=\alpha$, and $\ell(r)=0$ for all other rewards.
\item Let $u_j$ denote the vector of marginal utilities for $r_j$, i.e.
if $r_j=(a_1,\dots,a_n)$ then $u_j=(U_1(a_1),\dots,U_n(a_n))$.
Let $k$ denote the vector $(k_1,\dots,k_n)$.
With this notation, impose
\begin{subequations}
\label{eq:generalisedswingineq}
\begin{align}
\label{eq:generalisedswingineq1}
\forall j\in\{1,\dots,n-1\}\colon &&
(u_j-(1-\underline{\alpha}_j)u_0-\underline{\alpha}_j u_n)\cdot k&\ge 0 \\
\label{eq:generalisedswingineq2}
\forall j\in\{1,\dots,n-1\}\colon &&
(u_j-(1-\overline {\alpha}_j)u_0-\overline {\alpha}_j u_n)\cdot k&\le 0 \\
\label{eq:generalisedswingineq3}
&& 1\cdot k&=1
\end{align}
\end{subequations}
\end{enumerate}
The last constraint is simply another way of writing $\sum_{i=1}^n k_i=1$,
and fixes the multiplicative scaling of the joint utility function. These constraints define a convex set of weight vectors which represent the preferences of the expert. 

These assessments then lead to a set of linear inequalities
that determine a convex set of attribute weights.
In this decision analysis it is enough to consider extreme points of this set
(see \cref{tab:extremeweights}) to derive bounds on expected utility. 
The extreme points were here calculated using the double description method
\citep{1996:fukuda} through the rcdd package in R \citep{r}.
In our example, the number of extreme points is fairly limited.
For larger problems however, it might be required to use optimisation algorithms
that can work with the constraints directly.
Note that the number of extreme points
will depend in a non-trivial manner
on the elicited constraints represented by \cref{eq:generalisedswingineq},
and in particular on the number of attributes $n$,
and on the precise values  of $\underline{\alpha}_j$ and $\overline {\alpha}_j$.

\begin{table}
\caption{Extreme points of the convex set of attribute weights $(k_1, k_2, k_3, k_4)$ representing our incomplete preferences.
\label{tab:extremeweights}}
\begin{center}
\small
\begin{tabular}{l|c|rrrrrrrrr}
 Attribute & weight & 1 & 2 & 3 & 4 & 5 & 6 & 7 & 8 \\ 
 \hline
Biotic impact & $k_1$ & 0.37 & 0.36 & 0.39 & 0.39 & 0.26 & 0.25 & 0.28 & 0.27 \\ 
Longevity of impacts & $k_2$ & 0.31 & 0.30 & 0.26 & 0.26 & 0.36 & 0.36 & 0.31 & 0.31 \\ 
Feasibility & $k_3$ & 0.31 & 0.30 & 0.33 & 0.32 & 0.36 & 0.36 & 0.39 & 0.38 \\ 
Cost & $k_4$ & 0.01 & 0.03 & 0.01 & 0.03 & 0.01 & 0.04 & 0.02 & 0.04 \\ 
  \end{tabular}
\end{center}
\end{table}

\end{appendices}
\end{document}